\documentclass[final,onefignum,onetabnum]{siamart190516}

\usepackage{lipsum}
\usepackage{amsfonts}
\usepackage{graphicx}
\usepackage{epstopdf}
\usepackage{algorithmic}
\usepackage{mathrsfs}

\ifpdf
  \DeclareGraphicsExtensions{.eps,.pdf,.png,.jpg}
\else
  \DeclareGraphicsExtensions{.eps}
\fi

\newsiamremark{remark}{Remark}
\newsiamremark{hypothesis}{Hypothesis}
\crefname{hypothesis}{Hypothesis}{Hypotheses}
\newsiamthm{claim}{Claim}

\headers{Adjoint-based variational method}{S. Azimi, O. Ashtari, and T. M. Schneider}

\title{Adjoint-based variational method for constructing periodic orbits of high-dimensional chaotic systems\thanks{Submitted to the editors DATE.
\funding{This work was supported by the Swiss National Science Foundation (SNSF) under grant no. 200021-160088}}}

\author{Sajjad Azimi\thanks{Emergent Complexity in Physical Systems Laboratory (ECPS), \'Ecole Polythechnique F\'ed\'erale de Lausanne, CH-1015 Lausanne, Switzerland 
  (\url{https://ecps.epfl.ch}).}
\and Omid Ashtari\footnotemark[2]
\and Tobias M. Schneider\footnotemark[2] \thanks{\email{tobias.schneider@epfl.ch}}}

\usepackage{amsopn}

\ifpdf
\hypersetup{
  pdftitle={Adjoint-based variational method for constructing periodic orbits of high-dimensional chaotic systems},
  pdfauthor={S. Azimi, O. Ashtari, and T. M. Schneider}
}
\fi

\def\lp{\mathbf{l}}
\def\glp#1{\mathbf{#1}}
\def\ddr{\mathscr{L}}
\def\ddrv{\pmb{\mathscr{L}}}
\def\adj{\mathscr{L}^\dagger}
\def\adjv{\pmb{\mathscr{L}}^\dagger}
\def\lps{\mathscr{P}}
\def\res{r}
\def\resv{R}

\begin{document}

\maketitle

\begin{abstract}
    Chaotic dynamics in systems ranging from low-dimensional nonlinear differential equations to high-dimensional spatio-temporal systems including fluid turbulence is supported by non-chaotic, exactly recurring time-periodic solutions of the governing equations. These unstable periodic orbits capture key features of the turbulent dynamics and sufficiently large sets of orbits promise a framework to predict the statistics of the chaotic flow. Computing periodic orbits for high-dimensional spatio-temporally chaotic systems remains challenging as known methods either show poor convergence properties because they are based on time-marching of a chaotic system causing exponential error amplification; or they require constructing Jacobian matrices which is prohibitively expensive.
    We propose a new matrix-free method that is unaffected by exponential error amplification, is globally convergent and can be applied to high-dimensional systems. The adjoint-based variational method constructs an initial value problem in the space of closed loops such that periodic orbits are attracting fixed points for the loop-dynamics. We introduce the method for general autonomous systems. An implementation for the one-dimensional Kuramoto-Sivashinsky equation demonstrates the robust convergence of periodic orbits underlying spatio-temporal chaos. Convergence does not require accurate initial guesses and is independent of the period of the respective orbit.
\end{abstract}



\section{Introduction}
    Ideas from low-dimensional chaotic dynamical systems have recently led to new insights into high-dimensional spatio-temporally chaotic systems including fluid turbulence. The idea for a dynamical description of turbulence has a long history \cite{Thomas1942,Lin1944,Hopf1948} and stems from the observation that turbulent flows often show recognizable transient coherent patterns that recur over time and space \cite{Jimenez2018}. 
    Only in the last 15 years, however, has concrete progress allowed dynamical systems to be truly established as a new paradigm to study turbulence \cite{Kerswell2005,Eckhardt2007,Kawahara2012}. This progress is based on the discovery of unstable non-chaotic steady and time-periodic solutions of the fully nonlinear Navier-Stokes equations which leads to a description of turbulence as a walk through a connected forest of these dynamically connected invariant (`exact') solutions in the infinite-dimensional state space of the flow equations \cite{Gibson2008,Cvitanovic2010,Suri2017,Reetz2019a}. 
    
    Of special importance are time-periodic exactly recurring flows. These so-called unstable periodic orbits capture the evolving dynamics of the flow \cite{Kawahara2001} and form the elementary building blocks of the chaotic dynamics. Periodic orbits have been recognized as being key for understanding chaos since the 1880s \cite{Poincare1892,Ruelle1978,Gutzwiller1990}. Provided results from low-dimensional hyperbolic dissipative systems carry over to high-dimensional spatio-temporally chaotic systems, periodic orbits lie dense in the chaotic set supporting turbulence. The turbulent trajectory thus almost always shadows a periodic orbit. As a consequence, periodic orbit theory allows to express ergodic ensemble averages of the turbulent flow as weighted sums over periodic orbits. In these `cycle expansions', the statistical weight of an individual orbit is controlled by its stability features \cite{Auerbach1987,Cvitanovic1988, Artuso1990a, Artuso1990,Lan2010,chaosbook}. Sufficiently complete sets of periodic orbits for three-dimensional fluid flows may thus eventually allow to quantitatively describe statistical properties of turbulence in terms of exact invariant solutions of the underlying flow equations \cite{Chandler2013}. Even if a full description of turbulence in terms of periodic orbits remains beyond our reach, individual periodic orbits are of significant importance as they capture key physical processes underlying the turbulent dynamics and may inform control strategies \cite{Lasagna2017}. Consequently, robust tools for computing periodic orbits of high-dimensional spatio-temporally chaotic systems including three-dimensional fluid flows are needed. 
    
    High-dimensional spatio-temporal systems, including spectrally discretized three-dimensional fluid flow problems, are often characterized by more than $N=10^6$ highly coupled degrees of freedom.
    Computing periodic orbits of such high-dimensional strongly coupled systems remains computationally challenging. 
    The commonly used shooting method considers an initial value problem yielding trajectories satisfying the evolution equations and varies the initial condition until the solution closes on itself. To find the initial condition $u_0$ and the period $T$, Newton iteration is used to numerically solve the nonlinear equation $g(u_0,T) = f^T(u_0) - u_0$, where $f^T$ is the evolution of the state $u_0$ over time $T$. To solve this system of nonlinear coupled equations, a standard Newton method would require constructing the full Jacobian matrix with $\mathcal{O}(N^2)$ elements. This is practically impossible for high-dimensional strongly coupled systems with large $N$. 
    Key for computing periodic orbits of high-dimensional systems are thus \emph{matrix-free} Newton methods that do not construct the Jacobian matrix but only require successive evaluations of the function $g$, implying time-stepping of the evolution equations. Commonly used algorithms are Krylov subspace methods \cite{Kelley2003, Sanchez2004} including the Newton-GMRES-hook-step method by Viswanath \cite{Viswanath2007,Viswanath2009,Cvitanovic2010} as well as slight variations with alternative trust-region optimizations \cite{Dennis1996,Duguet2008a}.
    
    The matrix-free Newton approach is well suited for computing fixed points, where the `period' $T$ can be chosen arbitrarily, but the Newton approach poses fundamental challenges for periodic orbits. The defining property of a chaotic system is an exponential-in-time separation of trajectories which leads to a sensitive dependence on initial conditions. Very small changes in the initial condition $u_0$ are thus exponentially amplified by the required time-integration. Finding zeros of $g$ thus becomes an ill-conditioned problem. Consequently, an extremely good initial guess is required for the Newton method to converge. Generating sufficiently accurate initial guesses is very challenging and often impossible. 
    Owing to the finite numerical precision of double-precision arithmetic long and unstable orbits are even entirely impossible to converge.  
    Examples demonstrating the difficulty in finding periodic orbits of high-dimensional systems using shooting methods include the seminal work by Chandler and Kerswell \cite{Chandler2013}, who computed approximately $100$ orbits for a two-dimensional model flow and describe the time-consuming and tedious manual work to find initial guesses and trying to converge them. Likewise van Veen et al. \cite{VanVeen2018} recently computed a single periodic orbit for box turbulence with only moderate resolution of $64^3$ grid points. The authors reach a moderately small residual of $1.8\cdot10^{-4}$ and thus many orders of magnitude larger than machine precision only after ``several months of computing on modern GPU cards, due to the poor conditioning of the linear problems associated with Newton’s method". Consequently, more robust methods with larger radii of convergence than those of shooting methods are needed to compute periodic orbits of high-dimensional spatio-temporally chaotic systems. 
    
    For low-dimensional systems more robust methods for finding periodic orbits have been devised. Instead of starting from trajectories satisfying the evolution equations and varying the initial condition until the solution closes on itself, the variational approach suggested by Lan and Cvitanovi\'c \cite{Lan2004} reverses the approach: It starts from a closed loop in state space that does not satisfy the evolution equations and then adapts the loop until it solves the equations and a periodic orbit is found. To adapt the closed loop, the problem is recast as a minimization problem in the space of all closed loops. 
    The loop is driven towards a periodic orbit by minimizing a cost function that measures the deviation of the loop from an integral curve of the vector field induced by the governing equations. No time-marching along the orbit is required and the loop is adapted locally. Consequently, the variational method does not suffer from exponential error amplification and has a large radius of convergence. The robustness of the method has been demonstrated in the one-dimensional Kuramoto-Sivashinsky system \cite{Lan2008} for which Lasagna
     \cite{Lasagna2017} recently found more than $20\,000$ periodic orbits using $N=64$ Fourier modes to discretize the problem. 
     
    Unfortunately, the robust variational method of Lan and Cvitanovi\'c cannot be scaled to high-dimensional problems such as fluid turbulence. 
    The method is not matrix-free but requires the explicit construction of Jacobian matrices and their inversion. Moreover, accurate computations of tangents to the loop by finite differences require the loop to be represented by a sufficiently large number of closely-spaced instantaneous fields. The size of the Jacobian matrix to be inverted scales with the number of instantaneous fields $M$ and the spatial degrees of freedom $N$ as $\mathcal{O}(M^2 N^2)$.
    This scaling reflects the prohibitively large memory requirements for high-dimensional systems. The only attempt to apply the method to a higher-dimensional system we are aware of is Fazendairo et al. \cite{Fazendeiro2010, Boghosian2011} who study forced box-turbulence in a triple-periodic box using Lattice-Boltzmann computations. They provide evidence for the convergence of two periodic orbits but reaching a modestly small residual of $\mathcal{O}(10^{-5})$ on a relatively small $64^3$ spatial lattice requires tens of thousands of CPU cores. As stated by Fazendeiro et. al., even finding the shortest orbits of 3D flows using the method by Lan and Cvitanovi\'c requires petascale computing resources. Despite its robustness, the variational method by Lan and Cvitanovi\'c is thus too computationally expensive to be realistically used for high-dimensional spatio-temporally chaotic systems.
    
    Here we propose a novel matrix-free method that provides the same favorable convergence properties of the variational method by Lan and Cvitanovi\'c \cite{Lan2004,Lan2008} but can be applied to high-dimensional systems. 
    The method combines a variational approach similar to Lan and Cvitanovi\'c  with an adjoint-based minimization technique inspired by recent work of Farazmand \cite{Farazmand2016} on computing steady state solutions. Combining the variational approach with adjoints allows us to construct an initial value problem in the space of closed loops such that unstable periodic orbits become attracting fixed points of the dynamics in loop-space. Converging to a periodic orbit thus only requires evolving an initial guess under the dynamics in loop-space. 
    We develop the matrix-free adjoint-based variational method for general autonomous dynamical systems. As a proof-of-concept, the introduced method is applied to the one-dimensional Kuramoto-Sivashinsky equation (KSE) \cite{Kuramoto1976, Sivashinsky1977}. The KSE is a model system showing spatio-temporal chaos that has commonly been used as a sandbox model to develop algorithms that are eventually applied to three-dimensional fluid flows. We demonstrate the robust convergence of multiple periodic orbits of varying complexity and periods. The implementation utilizes a spectral Fourier discretization in the temporal direction to significantly reduce the prohibitively large memory requirements of the method by Lan and Cvitanovi\'c.
    
    The structure of the paper is as follows: First, the proposed method for computing periodic orbits is introduced for a general autonomous system. \Cref{sec:variational_method} describes the setup of the variational problem and \cref{sec:adjoint_formulation} discusses the adjoint-based minimization technique. In \cref{sec:KSE}, we apply the adjoint-based variational method to the KSE and demonstrate the convergence of periodic orbits in this spatio-temporally chaotic system. \Cref{sec:conclusion} summarizes the manuscript and discusses future applications to three-dimensional fluid turbulence.

\section{Variational method for finding periodic orbits}
\label{sec:variational_method}
    We consider a general dynamical system for an $n$-dimensional real field $\vec{u}$ defined over a spatial domain $\Omega\subset\mathbb{R}^d$ and varying in time $t$, 
    \begin{align*}
        \vec{u}:\Omega \times \mathbb{R} &\rightarrow \mathbb{R}^n,\\
        (\vec{x},t) &\mapsto \vec{u}(\vec{x},t).
    \end{align*}
    The evolution of the field $\vec{u}$ is first-order in time and governed by an autonomous partial differential equation (PDE) of the form
    \begin{equation}
        \label{eq:dynamical_system_PDE}
        \frac{\partial \vec{u}}{\partial t}=\mathcal{N}(\vec{u}).
    \end{equation}
    The nonlinear differential operator $\mathcal{N}$ enforces boundary conditions at $\partial\Omega$, the boundaries of the spatial domain $\Omega$. 
    A periodic orbit is a temporally periodic solution of the governing equation,
    \begin{equation}\label{eq:PO_def}
        f^T(\vec{u})-\vec{u}=\vec{0},
    \end{equation}
    where $f^T = \int_{t}^{t+T} \mathcal{N} dt'$ indicates the nonlinear evolution over the period $T$.
    
    The shooting method considers solutions of the initial value problem and varies the initial condition $\vec{u}_0(\vec{x})$ until the solution closes on itself and becomes periodic. \Cref{eq:PO_def} is thus treated as an algebraic equation for the initial condition and the period. An alternative approach is to consider already time-periodic fields and vary those until they satisfy the governing equations. Instead of identifying an initial condition as in a shooting method, we consider the entire orbit as a solution of a boundary value problem in the $(d+1)$-dimensional space-time domain. To ensure periodicity of the solution in time, the boundary conditions in space are augmented by periodic boundary conditions in time. The field $\vec{u}(\vec{x},t)$ is thus defined on $\Omega \times [0,T)_\text{periodic}$.
    
    The length of the domain in time $T$ is unknown and needs to be determined as part of the solution. To convert the problem to a boundary value problem on a fixed domain, we rescale time $t \mapsto s:=t/T$, where $s$ denotes the normalized time coordinate. The rescaled field
    \begin{equation*}
        \vec{\tilde{u}}(\vec{x},s) := \vec{u}(\vec{x},s\cdot T),
    \end{equation*}
    is defined on a fixed domain
    \begin{align*}
        \vec{\tilde{u}}:\Omega \times [0,1)&_\text{periodic} \rightarrow \mathbb{R}^n,\\
        (\vec{x},s)& \mapsto \vec{\tilde{u}}(\vec{x},s).
    \end{align*}
    A periodic orbit is characterized by the space-time field $\vec{\tilde{u}}(\vec{x},s)$ and the period $T$ satisfying 
    \begin{equation}
        \label{eq:BVP_PDE}
        -\frac{1}{T}\frac{\partial \vec{\tilde{u}}}{\partial s}+\mathcal{N}(\vec{\tilde{u}})=\vec{0}.
    \end{equation}
    Boundary conditions in space remain unchanged with respect to the dynamical system \cref{eq:dynamical_system_PDE} and are complemented by periodic boundary conditions in the temporal direction $s$. To simplify the notation, the overhead tilde is omitted in the remainder of the article. 
    
    A periodic orbit is defined by the combination of a field $\vec{u}(\vec{x},s)$ and a period $T$ that together satisfy the boundary value problem \cref{eq:BVP_PDE}. Geometrically the periodic orbit is a closed trajectory in state space. To characterize general closed curves in state space, we define a \emph{loop} $\lp(\vec{x},s)$ as a tuple of a field $\vec{u}(\vec{x},s)$ and a period $T$. A loop does not necessarily satisfy the PDE of the boundary value problem \cref{eq:BVP_PDE} but shares all boundary conditions in space and time with periodic orbits. We denote the space of all loops by 
    \begin{align}
        \label{eq:loop_def}
        \lps=
        \left\{\lp(\vec{x},s)=
        \begin{bmatrix} 
            \vec{u}(\vec{x},s)\\
            T
        \end{bmatrix}\;
        \Bigg|
        \begin{matrix} 
            \vec{u}:\Omega \times [0,1)_\text{periodic} \rightarrow \mathbb{R}^n,\; T \in \mathbb{R}^+ \\
            \vec{u}\ \text{satisfies BC at }\partial\Omega 
            \text{ and is periodic in}\ s
        \end{matrix}
        \right\}.    
    \end{align}
    Periodic orbits are specific elements of the loop-space $\lps$ that satisfy the PDE \cref{eq:BVP_PDE}. A general loop only satisfies the boundary conditions but not the PDE. 
    
    The idea of the variational method is to consider an initial loop $\lp_0(\vec{x},s)\in\lps$ and to evolve the loop until it satisfies the boundary value problem \cref{eq:BVP_PDE}. The loop thereby converges to a periodic orbit. To evolve a loop towards a periodic orbit we minimize the cost function $J$ measuring the deviation of a loop from a solution of the boundary value problem,
    \begin{align}
        \label{eq:cost_function}
        \begin{split}
            J:\lps &\rightarrow \mathbb{R}^+,\\
            \lp \mapsto J(\textbf{l}) &:=\int_0^1\int_\Omega \vec{\res}\cdot \vec{\res} d\vec{x}ds.
        \end{split}
    \end{align}
    where $\vec{\res}$ is the residual of \Cref{eq:BVP_PDE}:
    \begin{equation}
        \label{eq:res_definition}
        \vec{\res}=-\frac{1}{T}\frac{\partial \vec{u}}{\partial s}+\mathcal{N}(\vec{u}).
    \end{equation}
    The cost function $J$ penalizes a nonzero residual $\vec{\res}$.
    For a periodic orbit $J$ is zero otherwise it takes positive values. Thus, absolute minima of $J$ correspond to periodic orbits. The problem of finding periodic orbits has thereby been converted into an optimization over loop-space $\lps$.
    
    \begin{figure}
        \centering
        $(a)$
        \includegraphics[width=0.35 \linewidth]{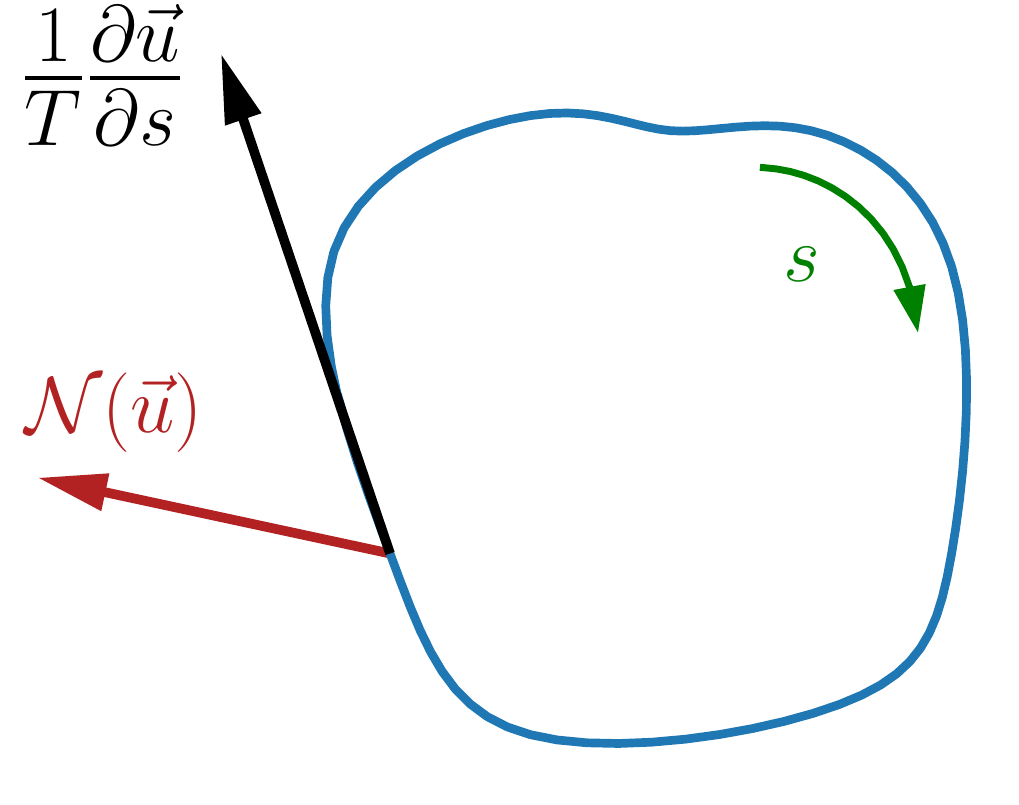}\\
        $(b)$
        \includegraphics[width=0.85\linewidth]{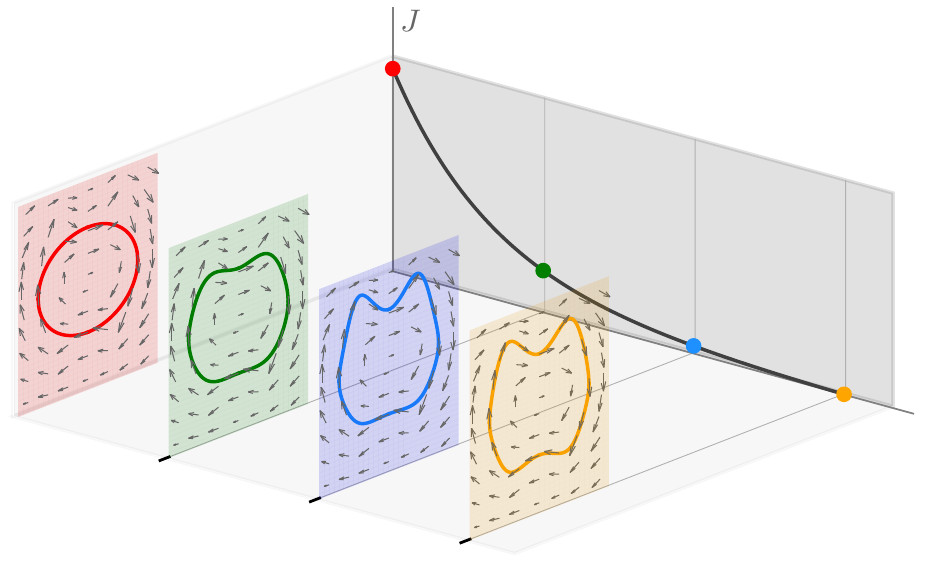}
        \caption{Schematic of the variational method for finding periodic orbits. (a) An arbitrary closed loop (blue line) parametrized by $s \in[0,1)$ does not satisfy the governing equations as its loop tangent $\partial \vec{u}/\partial t = T^{-1}\partial \vec{u}/\partial s$ is misaligned relative to the vector field $\mathcal{N}(\vec{u})$ induced by the dynamical system. (b) Minimizing a cost function $J$ measuring the misalignment between the vector field and the loop tangent deforms the loop. When the global minimum of the cost function with $J=0$ is reached the tangent vectors everywhere match the flow, $\partial \vec{u}/\partial t = \mathcal{N}(\vec{u})$. The loop becomes an integral curve of the vector field and a periodic orbit is identified.}
        \label{fig:variational_idea}
    \end{figure}
    
    Geometrically, minimizing the cost function corresponds to deforming a closed curve, a loop, in the system's state space, the space spanned by all instantaneous fields $\vec{u}(\vec{x})$ satisfying the boundary conditions, until the loop becomes an integral curve of the vector field $\mathcal{N}(\vec{u})$ induced by the dynamical system. The loop thereby becomes a solution of the PDE and represents a periodic orbit. This is schematically shown in \Cref{fig:variational_idea}. At each point $\vec{u}$ along the loop, the vector field defines the flow direction $\mathcal{N}(\vec{u})$ while $\partial \vec{u}/\partial t = T^{-1}\partial \vec{u}/\partial s$ is the tangent vector along the loop (see panel $a$). The cost function $J$ measures the misalignment between the vector field and the loop's tangent vectors integrated along the entire loop. Consequently, minimizing $J$ towards its absolute minimum $J=0$ deforms the loop until the tangent vectors everywhere match the flow and the loop becomes an integral curve of the vector field, as exemplified in panel $b$. The loop is locally deformed to align with the vector field and no time-marching causing exponential instabilities is required.
    
\section{Adjoint-based method for minimizing the cost function $J$}
\label{sec:adjoint_formulation}
    We recast the problem of finding periodic orbits as a minimization problem in the space of all loops. Absolute minima of the cost function $J$ with value $J=0$ correspond to periodic orbits. To minimize $J$ without constructing Jacobians we develop an adjoint-based approach inspired by the recently introduced method by Farazmand \cite{Farazmand2016} who computes equilibria of a two-dimensional flow. We construct an initial value problem in loop-space $\lps$ whose dynamics monotonically decreases the cost function $J$ until a minimum of $J$ is reached.
    
    To derive an appropriate variational dynamics in loop-space, we define the space of generalized loops:
    \begin{align}
        \label{eq:gloop_def}
        \lps_g=
        \left\{\glp{q}(\vec{x},s)=
        \begin{bmatrix} 
            \vec{q_1}(\vec{x},s)\\
            q_2
        \end{bmatrix}\;
        \Bigg|
        \begin{matrix} 
            \vec{q_1}:\Omega \times [0,1)_\text{periodic} \rightarrow \mathbb{R}^n,\; q_2 \in \mathbb{R} \\
            \vec{q_1}\ \text{is periodic in}\ s
        \end{matrix}
        \right\}.    
    \end{align}
    Elements $\glp{q} \in \lps_g$ do not necessarily satisfy the spatial boundary condition of periodic orbits at $\partial\Omega$ and are thus termed generalized loops. Obviously, the space of loops $\lps$ is a subset of the space of generalized loops $\lps \subset \lps_g$. For a loop, the components of the generalized loop have specific meaning, $\vec{q_1}=\vec{u}$ and $q_2=T$. Throughout this paper, generalized loops are denoted by boldface letters.
    The space of generalized loops $\lps_g$ carries a real-valued inner product 
    \begin{align}
        \label{eq:inner_product_definition}
        \begin{split}
        \left<\ ,\ \right>:\lps_g &\times \lps_g \rightarrow \mathbb{R},\\
        \left<\glp{q},\glp{q}'\right>=
        \left<\;
        \begin{bmatrix}
            \vec{q_1}\\
            q_2
        \end{bmatrix},
        \begin{bmatrix}
            \vec{q'_1}\\
            q'_2
        \end{bmatrix}
        \;\right> &=\int_0^1\int_\Omega{\vec{q_1} \cdot \vec{q'_1}}d\Vec{x}ds+q_2q'_2,
        \end{split}
    \end{align}
    and an $L_2$-norm
    \begin{equation}
    \label{eq:L2norm_definition}
        ||\glp{q}||=\sqrt{\left<\glp{q},\glp{q}\right>}=\sqrt{\int_0^1\int_\Omega{\vec{q_1} \cdot \vec{q_1}}d\vec{x}ds+q^2_2}.
    \end{equation}
    
    The objective is to construct a dynamical system in the space of loops $\lps$ such that along its solutions the cost function $J$ monotonically decreases and periodic orbits become attracting fixed points of the dynamical system. We parametrize the evolution of loops in $\lps$ by a fictitious time $\tau$: $\lp(\tau)=[\vec{u}(\vec{x},s;\tau); T(\tau)]$ and define an evolution equation,
    \begin{equation}
    \label{eq:lp_PDE}
        \frac{\partial \lp}{\partial \tau}=\glp{G(\lp)}
    \end{equation}
    with operator $\glp{G}$ chosen such that
    \begin{equation}
    \label{eq:objective_revisited}
        \frac{\partial J}{\partial \tau}\leq0\quad\forall\ \tau.
    \end{equation}
    
    The rate of change of $J$ along solutions of \Cref{eq:lp_PDE} is (see \Cref{sec:d(J)/d(tau)} for details)
    \begin{equation}
    \label{eq:rate_of_J}
        \frac{\partial J}{\partial \tau}=2\left<\ddrv(\lp;\glp{G}),\glp{\resv}\right>.
    \end{equation}
    where $\glp{\resv}\in\lps_g$ is a generalized loop
    \begin{equation}
    \label{eq:res_v_definition}
        \glp{\resv}(\lp)=
        \begin{bmatrix}
            \vec{r}\\
            0
        \end{bmatrix},
    \end{equation}
    with $\vec{r}(\textbf{l})$ the residual field \cref{eq:res_definition}.
    $\ddrv(\lp;\glp{G})$ is the directional derivative of the residual $\glp{\resv}$ in the direction $\glp{G}$, evaluated for the current loop $\lp$:
    \begin{equation}
    \label{eq:directional_derivative}
        \ddrv(\lp;\glp{G})=\lim_{\epsilon\to 0}{\frac{\glp{\resv}(\lp+\epsilon \glp{G})-\glp{\resv}(\lp)}{\epsilon}}.
    \end{equation}
    Using the adjoint of the directional derivative, we express \Cref{eq:rate_of_J} as
    \begin{equation}\label{eq:rate_of_J_adjoint}
        \frac{\partial J}{\partial \tau}=2\left<\glp{G},\adjv(\lp;\glp{\resv})\right>
    \end{equation}
    where $\adjv$ is the adjoint operator of $\ddrv$ with
    \begin{equation}
    \label{eq:adjoint_definition}
        \left<\ddrv(\glp{q};\glp{q}'),\glp{q}''\right>=\left<\glp{q}',\adjv(\glp{q};\glp{q''})\right>,
    \end{equation}
    for all generalized loops $\glp{q}$, $\glp{q}'$ and $\glp{q}''$.
    This form allows to enforce the monotonic decrease of the cost function $J$ by explicitly choosing the operator $\glp{G}$:
    \begin{equation}
    \label{eq:G_as_adjoint}
        \glp{G}=-\adjv(\lp;\glp{\resv}).
    \end{equation}
    With this choice for $\glp{G}$, the cost function evolves as
    \begin{equation}
        \label{eq:J_variation}
        \frac{\partial J}{\partial \tau}=2\left<-\adjv(\lp;\glp{\resv}),\adjv(\lp;\glp{\resv})\right>=-2\left|\left|\adjv(\lp;\glp{\resv})\right|\right|^2\leq0.
    \end{equation}
    Thus, along solutions of $\partial \lp/\partial \tau = \glp{G}(\lp)=-\adjv(\lp;\glp{\resv})$ the cost function $J$ is guaranteed to monotonically decrease. 
    
    To find a periodic orbit using the adjoint approach, an initial loop is advanced under the dynamical system in loop-space, until a minimum of the cost function, corresponding to an attracting fixed point with $\partial_\tau \lp = \glp{0}$, is reached. If an absolute minimum, $J=0$, is reached, the loop satisfies the boundary value problem \cref{eq:BVP_PDE} and represents a periodic orbit.  The cost function $J$ is invariant under a reparametrization $s \mapsto s' = (s + \sigma) \bmod{1}$ corresponding to a phase shift by $\sigma$ in the temporal periodic direction. Consequently, the phase of the minimizing loop is not chosen by the adjoint-based variational method but depends on the initial condition.
    
\section{Application to Kuramoto-Sivashinsky equation}
\label{sec:KSE}
    We demonstrate the adjoint-based variational method for the one-dimensional Kuramoto-Sivashinsky equation (KSE) \cite{Kuramoto1976, Sivashinsky1977}. This nonlinear partial differential equation for a one-dimensional field $u(x,t)$ on a 1D periodic interval $x \in [0,L)=\Omega$ reads
    \begin{equation}
        \label{eq:KS}
        \frac{\partial u}{\partial t}=-u\frac{\partial u}{\partial x}-\frac{\partial^2 u}{\partial x^2}-\nu\frac{\partial^4 u}{\partial x^4}\;;\quad\quad x\in[0,L)_\text{periodic},\;t\in\mathbb{R}
    \end{equation}
    with a constant 'superviscosity' $\nu>0$. The KSE has the general form of \Cref{eq:dynamical_system_PDE} with $n=d=1$. We denote the scalar spatial coordinate by $x$. Rescaling the field $u$ by the inverse of $L$ indicates that the only control parameter is $\mathbb{L}=L/\sqrt{\nu}$ the ratio of the domain length and the square-root of the constant $\nu$. Consequently, fixing the domain length $L$ and varying $\nu$ is equivalent to fixing $\nu$ and treating $L$ as a control parameter. Both scalings are used in literature. Here, we fix $\nu=1$ and consider $L$ as the control parameter. The equivariance group of the KSE contains continuous shifts in $x$ and the discrete center symmetry,
    \begin{equation}
        \label{eq:KS_sym}
        x\rightarrow-x\;;\quad u\rightarrow-u.
    \end{equation}
    We discuss periodic orbits both in the full unconstrained space and in the subspace of fields invariant under the discrete center symmetry. 
    
    The trivial solution of the KSE, $u=\text{const}$, is linearly unstable for $L>2\pi\sqrt{\nu}$ \cite{Cvitanovic2010a}. A series of bifurcations leads to increasingly complex dynamics when $L$ is increased. We consider the parameter value $L=39$ where the KSE shows spatio-temporally chaotic dynamics reminiscent of turbulence \cite{Smyrlis1996}.
    
\subsection{Formulation of the adjoint-based method for the KSE}
    For the 1D-KSE a loop consists of a one-dimensional field $u(x,s)$ defined over $[0,L) \times [0,1)$ and the period $T$. The residual of the boundary value problem for a periodic orbit \cref{eq:res_definition}, expressed as generalized loop $\glp{\resv}$ (see \Cref{eq:res_v_definition}), is
    \begin{equation}
        \label{eq:KS_R}
        \renewcommand\arraystretch{1.8}
        \glp{\resv}(\mathbf{l})=
        \begin{bmatrix}
            \res(\mathbf{l})\\
            0
        \end{bmatrix}=
        \begin{bmatrix}
            -\dfrac{1}{T}\dfrac{\partial u}{\partial s}-u\dfrac{\partial u}{\partial x}-\dfrac{\partial^2 u}{\partial x^2}-\dfrac{\partial^4 u}{\partial x^4}\\
            0
        \end{bmatrix},
    \end{equation}
    where vector notation has been suppressed because the dimension of the field is $n=1$.
    
    The dynamical system in loop-space for which the cost function monotonically decreases and periodic orbits become attracting fixed points is based on the adjoint operator of the directional derivative of $\glp{\resv}$. Partial integration directly yields the adjoint operator for the KSE problem (see \Cref{sec:K-S_adjoint_operator}),
    \begin{equation}
        \label{eq:KS_L_dagger}
        \renewcommand\arraystretch{2.5}
        \pmb{\mathscr{L}}^\dagger(\lp;\glp{\resv})=
        \begin{bmatrix}
            \dfrac{1}{T}\dfrac{\partial \res}{\partial s}+u\dfrac{\partial \res}{\partial x}-\dfrac{\partial^2 \res}{\partial x^2}-\dfrac{\partial^4 \res}{\partial x^4}\\
            \displaystyle\int_0^1\int_0^L{\frac{1}{T^2}\frac{\partial u}{\partial s}\res}dxds
        \end{bmatrix}.
    \end{equation}
    Consequently, the dynamical system in loop-space $\partial\lp/\partial\tau=-\pmb{\mathscr{L}}^\dagger(\lp;\glp{\resv})$ (see \cref{eq:G_as_adjoint}) minimizing the cost function $J$ is
    \begin{equation}
        \label{eq:KS_G}
        \renewcommand\arraystretch{2.5}
        \frac{\partial \lp}{\partial \tau}=
        \begin{bmatrix}
            \dfrac{\partial u}{\partial \tau}\\
            \dfrac{\partial T}{\partial \tau}
        \end{bmatrix}=
        \begin{bmatrix}
            -\dfrac{1}{T}\dfrac{\partial \res}{\partial s}-u\dfrac{\partial \res}{\partial x}+\dfrac{\partial^2 \res}{\partial x^2}+\dfrac{\partial^4 \res}{\partial x^4}\\
            -\displaystyle\int_0^1\int_0^L{\frac{1}{T^2}\frac{\partial u}{\partial s}\res}dxds
        \end{bmatrix}.
    \end{equation}
    The first component of \Cref{eq:KS_G} prescribes the deformation of the field $u(x,s)$, while the second component updates the period $T$. 
    
    The dynamical system in loop-space formulated for the KSE, \Cref{eq:KS_G}, is equivariant with respect to the discrete symmetry:
    \begin{equation}
        \label{eq:KS_loop_sym}
        \Xi:\;(x,s)\rightarrow(-x,s)\;;\quad
        \begin{bmatrix}
            u\\
            T
        \end{bmatrix} 
        \rightarrow
        \begin{bmatrix}
            -u\\
            T
        \end{bmatrix}
        .
    \end{equation}
    If an initial loop is invariant under the action of $\Xi$, the evolution in $\tau$ will preserve the symmetry. Since the transformation of the instantaneous field $x\rightarrow -x; u(\cdot, s) \rightarrow -u(\cdot, s)$ for all $s \in [0,1)$ corresponds to the center-symmetry \cref{eq:KS_sym} of the KSE equation, the dynamical system in loop-space also preserves the center symmetry of the KSE. An initial loop with field component within the center-symmetric subspace of KSE is invariant under $\Xi$, which is preserved under $\tau$-evolution. Consequently, the adjoint-based variational method preserves the discrete center-symmetry of the KSE.

\subsection{Numerical implementation}
    Expressing the field component of the dynamical system \cref{eq:KS_G} in terms of $u$ using \Cref{eq:KS_R} yields,
    \begin{equation}
        \label{eq:KS_G_1}
        \frac{\partial u}{\partial \tau} =  G_{1,\text{L}} + G_{1,\text{NL}},
    \end{equation}
    where the linear and nonlinear terms have the form,
    \begin{align*}
        G_{1,\text{L}} = \frac{1}{T^2}\frac{\partial^2 u}{\partial s^2} - \frac{\partial^8 u}{\partial x^8}-& 2\frac{\partial^6 u}{\partial x^6}-\frac{\partial^4 u}{\partial x^4}\\
        G_{1,\text{NL}} = -5\frac{\partial^4 u}{\partial x^4}\frac{\partial u}{\partial x} - 10\frac{\partial^3 u}{\partial x^3}\frac{\partial^2 u}{\partial x^2} - 3\frac{\partial^2 u}{\partial x^2}\frac{\partial u}{\partial x} +& u^2\frac{\partial^2 u}{\partial x^2} +  u\left(\frac{\partial u}{\partial x}\right)^2+\frac{2u}{T}\frac{\partial^2 u}{\partial x\partial s}+\frac{1}{T}\frac{\partial u}{\partial x}\frac{\partial u}{\partial s}.
    \end{align*}
    The field $u(x,s)$ is defined on a doubly-periodic space-time domain.
    We thus numerically solve the evolution equation with a pseudospectral method \cite{Canuto2006} using a Fourier discretization in both space and time. The spectral representation with $M$ modes in space and $N$ modes along the temporal direction is,
    \begin{align}
        u(x_m,s_n)=\sum_{j=-\frac{M}{2}}^{\frac{M}{2}-1}\;\sum_{k=-\frac{N}{2}}^{\frac{N}{2}-1}{\hat{u}_{j,k}\exp{\left\{2\pi i\left(\frac{mj}{M}+\frac{nk}{N}\right)\right\}}}.
        \label{eq:2D_Fourier}
    \end{align}
    In physical space, the field is represented by grid values at the Gauss-Lobatto collocation points $\{u(x_m,s_n)\}$ with $(x_m,s_n)=(mL/M,n/N)$ and index ranges $0\le m \le M-1$ and $0\le n \le N-1$. In spectral space, the set of discrete Fourier coefficients $\{\hat{u}_{j,k}\}$ with $-M/2\le j \le M/2-1$ and $-N/2\le k \le N/2-1$ represents the field.  
    In spectral space, the evolution equation \cref{eq:KS_G_1} for each Fourier coefficient of the field takes the form
    \begin{equation}
        \frac{\partial \hat{u}_{j,k}}{\partial \tau}=\left[-\left(\frac{2\pi k}{T}\right)^2-\left(\frac{2\pi j}{L}\right)^8+2\left(\frac{2\pi j}{L}\right)^6-\left(\frac{2\pi j}{L}\right)^4\right]\hat{u}_{j,k}+(\hat{G}_{1,\text{NL}})_{j,k}\;,
    \end{equation}
    where the discrete Fourier transform is indicated by a hat. To evaluate the nonlinear term $\hat{G}_{1,\text{NL}}$ derivatives are calculated in spectral space and transformed to physical space, where products are pointwise operations. Transforming the result back to spectral space yields the required terms. In both the spatial and temporal direction dealiasing following the 2/3 rule \cite{Canuto2006} is applied.
    To advance the evolution equation \cref{eq:KS_G_1} in the fictitious time $\tau$ we implement a semi-implicit time-stepping method. An implicit-explicit Euler method treats the linear terms implicitly and the nonlinear terms $\hat{G}_{1,\text{NL}}$ are discretized explicitly. 
    
    The second component of the evolution equation \cref{eq:KS_G} evolves the period of the loop $T$. We use an explicit Euler method for time-stepping. The integral defining the right-hand-side is evaluated analogous to the pseudo-spectral treatment of the nonlinear terms in the evolution equation of the field. The integrand is evaluated in physical space followed by transformation to spectral space, where the integral is given by the $(0,0)$ Fourier mode multiplied by $L$. 
    
    Since the purpose of defining the initial value problem in loop-space is to identify attractors corresponding to solutions of the boundary value problem for periodic orbits, stability and simplicity of the implementation are more important than accuracy when choosing a time-stepping scheme. The simple Euler method is only first order accurate in $\tau$ but remains stable for the chosen fixed time step $\Delta \tau = 0.15$.
    
\subsection{Initial guesses and convergence to periodic orbits}
\label{sec:initial_converge}
    The adjoint-based variational method advances some initial loop under the dynamical system that minimizes the cost function $J$.  If a minimum  with $J=0$ is reached the loop satisfies the boundary value problem for a periodic orbit. Initial guesses for the procedure are extracted from chaotic solutions of the KSE \cref{eq:KS} $u(x,t)$. The common approach for generating guesses used in conjunction with Newton-GMRES-based shooting methods extracts close recurrences measured in terms of the $L_2$-distance from minima of the recurrence map $c(t,T) = ||u(\cdot, t + T) -  u(\cdot,t)||$ \cite{Auerbach1987}. Here, the $L_2$-norm is given by      
    \begin{align}\label{eq:conventinal_L2norm}
        ||u||(t)=\sqrt{\int_0^L{u(x,t)^2}dx}.
    \end{align}
    Exploiting the large radius of convergence of the variational method, we here choose a much simpler and computationally significantly cheaper method. Initial guesses are extracted from close recurrences in a one-dimensional projection of the solution. Specifically, we consider subsequent maxima in the time series of $||u||(t)$ where $||u||(t+T) \approx ||u||(t)$. The segment of the solution between those subsequent maxima yields the field component of the initial loop. To ensure a smooth closed loop with field component satisfying periodic boundary conditions in the temporal direction, the solution segment is Fourier-transformed in time and high-frequency components are filtered out \cite{Lan2004}. The double-periodic field $u_0(x,s)$ complemented by the period defines an initial guess $\lp_0 = [u_0(x,s);T]$. 
    
    The initial guess $\lp_0$ is evolved under the dynamical system in loop-space \cref{eq:lp_PDE}. Along the evolution the cost function $J$ is guaranteed to monotonically decrease and reach a minimum. Consequently, the adjoint-based variational method is globally convergent. However, it is not guaranteed that an absolute minimum with $J=0$ is reached but the dynamics may asymptote towards a local minimum with $J>0$. If a global minimum is reached, a periodic orbit satisfying the boundary value problem \cref{eq:BVP_PDE} is found. We consider a periodic orbit converged, when $\sqrt{J} < 10^{-12}$ is achieved. The periodic orbit corresponds to an attracting fixed point of the dynamical system in loop-space so that we expect exponential convergence at a rate controlled by the leading eigenvalue of the loop dynamics linearized around the attracting fixed point. 
    
\subsection{Results and discussion}
\label{sec:results}
    \begin{figure}
        \centering
        \includegraphics[width=1.0 \linewidth]{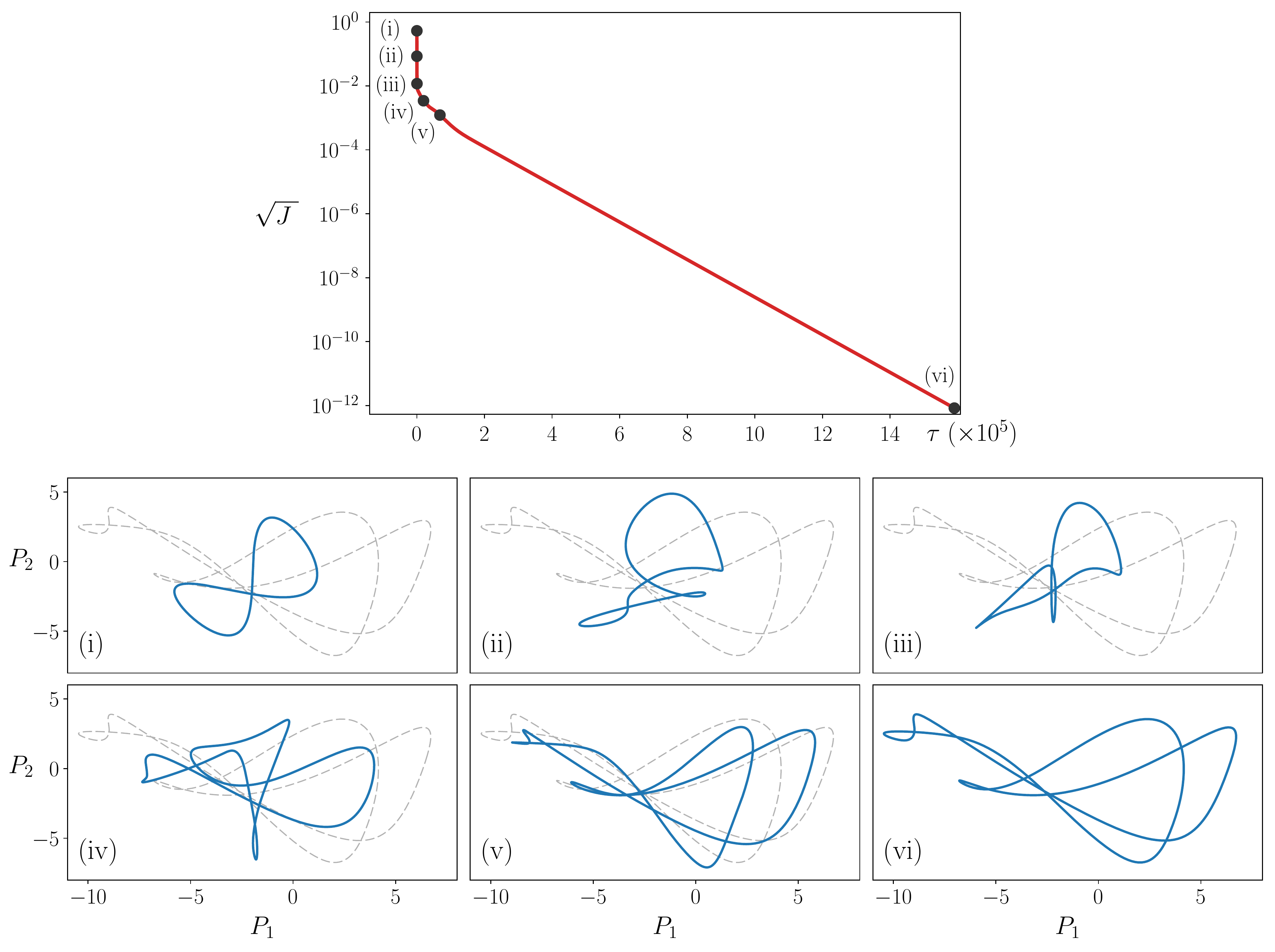}
        \caption{Convergence of the adjoint-based variational method for finding periodic orbits of the KSE: The initial value problem in loop-space evolves loops such that the cost function J decreases monotonically along the fictitious time $\tau$ (top). The exponential decay of J towards zero indicates convergence towards a periodic orbit satisfying $J=0$. Geometrically, the variational dynamics deforms a closed loop until it becomes an integral curve of the flow and thus a periodic orbit of the KSE. This is shown in the bottom panel, where the evolution of the loop is visualized in a two-dimensional projection of state space. Blue solid lines indicate the evolving loop at times indicated in the top panel. The dashed gray line is the converged periodic orbit.
        The state space projections $P_1(s)$ and $P_2(s)$ are defined by the imaginary parts of the first and second spatial Fourier coefficients of the field $u(x)$. 
        }
        \label{fig:Norm_Loops}
    \end{figure}
    
    We demonstrate the adjoint-based variational method to construct periodic orbits of the KSE for the parameter value $L=39$. At this value, the dynamics is chaotic and a large number of unstable periodic orbits are known to exists \cite{Lasagna2017}. 
    Periodic orbits of the KSE are found by evolving initial loops under the dynamical system in loop-space \cref{eq:KS_G}. The pseudo-spectral method uses $64 \times 64$ Fourier modes in spatial and temporal directions to discretize the field $u(x,s)$. A fixed time step of $\Delta \tau = 0.15$ leads to stable time-stepping. 
    
    Periodic orbits of the KSE are attracting solutions of an initial value problem in the space of loops $\lps$ that monotonically decreases the cost function $J$, as shown in \Cref{fig:Norm_Loops}. In the top panel, the square root of the cost function, $\sqrt{J}$, as a function of the fictitious time $\tau$ is shown. After $\tau\approx 1.5\cdot 10^6$ the convergence criterion $\sqrt{J}\le 10^{-12}$ is reached. Since the cost function $J$ is the average of $\int_\Omega r^2 dx$ over $s$, the square root of $J$ scales with the $L_2$-norm \cref{eq:conventinal_L2norm} of the residual field $r$ and should be used as the convergence criterion. Along the evolution of the loop with $\tau$ the cost function $J$ monotonically decreases. After an initial fast decrease, $\sqrt{J}$ decays exponentially with $\tau$. This suggests the convergence towards a periodic orbit along the leading eigendirection of the dynamical system in loop-space linearized about the attracting fixed point. Geometrically, the dynamical system in loop-space \cref{eq:KS_G} continuously deforms the initial loop until the loop satisfies the KSE and thereby becomes a periodic orbit. The deformation is visualized in the bottom panel, where the evolution of the loop shown in a two dimensional projection of the state space. A very substantial deformation of the loop is associated with the fast decrease of $J$ within the initial $10\%$ of the integration time. 
    
    \begin{figure}
        \centering
        \includegraphics[width=0.65\textwidth]{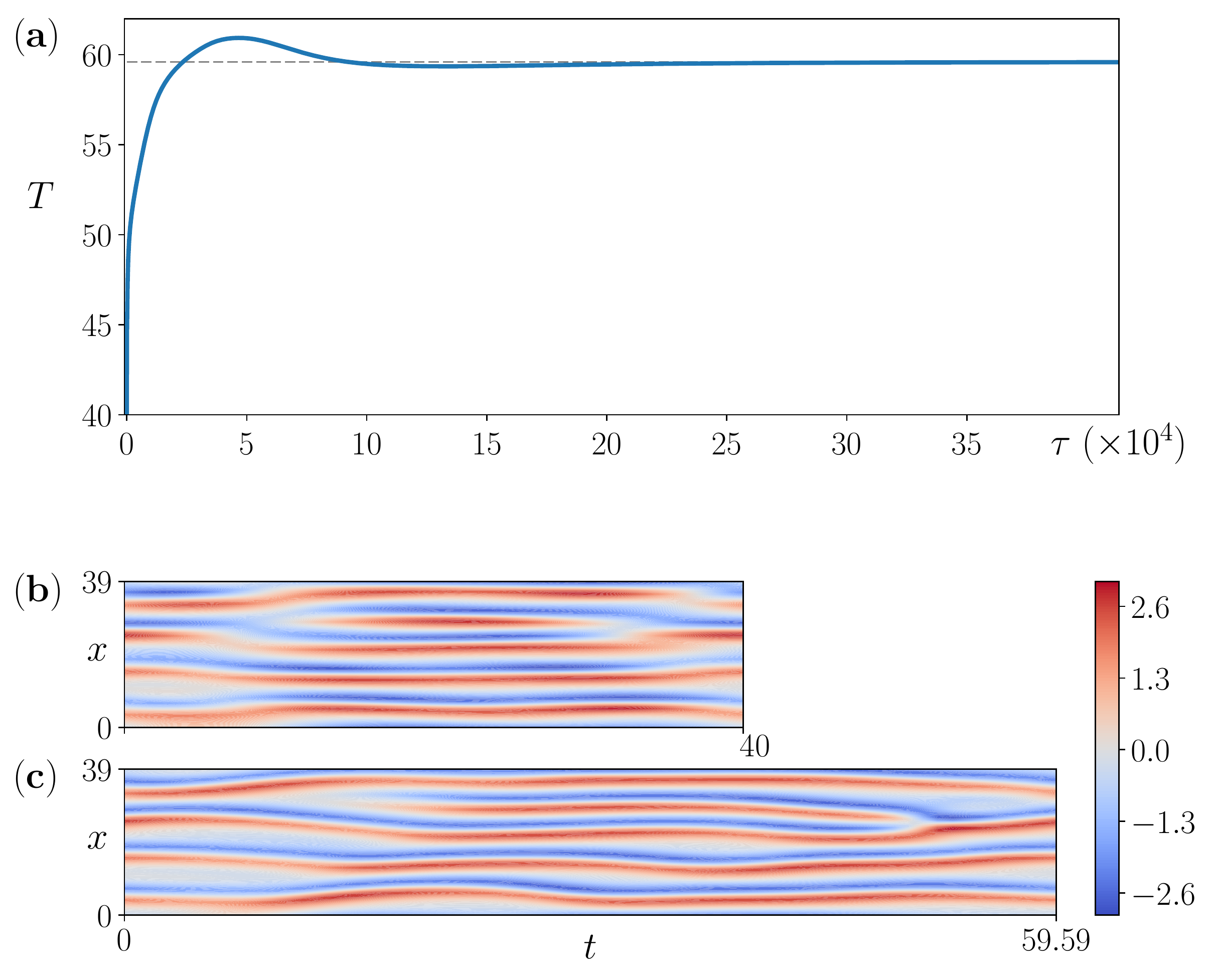}
        \caption{A periodic orbit is characterized by the combination of the field $u(x,s)$ on a fixed double-periodic space-time domain and the time period $T$ that rescales the temporal direction $s \rightarrow t=T\cdot s$. The variational dynamics adapts $T$ until the period of the periodic orbit is determined (top). Finding the period $T$ corresponds to determining the length of the domain in time $t$. This is evidenced by space-time contours of the solution $u(x,t=T\cdot s)$ for the initial condition (b) and the converged periodic orbit (c). The period of the initial loop and the periodic orbit are $T=40$ and $T=59.59$, respectively. }
        \label{fig:T_Contour}
    \end{figure}
    
    In addition to the two-dimensional field defined over the fixed space-time domain $[0,L)\times[0,1)$, the corresponding period $T$ is required to define a loop. Evolving a loop towards a periodic orbit implies finding the period $T$, which re-scales the temporal length of the space-time domain $s\rightarrow t = T\cdot s$ and thereby determines the length of extension of the domain in the direction of time $t$. \Cref{fig:T_Contour} shows the convergence of $T$ to the period of the periodic orbit together with the space-time contours of the corresponding initial loop $u_{0}(x,t=T\cdot s)$  and the converged periodic orbit $u(x,t=T\cdot s)$.  As for the geometry of the loop (\Cref{fig:Norm_Loops}) substantial changes in the period $T$ under the adjoint-based variational dynamics are mostly observed within the initial $10\%$ of the integration of the dynamical system in loop-space \cref{eq:KS_G}. Already at $\tau=2 \cdot 10^5$, $T$ is very close to the period of the periodic orbit $T=59.59$. We omit data beyond $\tau=4 \cdot 10^5$ from \Cref{fig:T_Contour} since changes would not be visible. 
    
    The fast initial decrease of the cost function $J$ followed by a slow exponential decay towards zero suggests that the loop approaches the periodic orbit along the leading eigendirection of the loop dynamics linearized around the attracting fixed point. Most of the computational efforts are spent on following the exponential decay until the cost function has reached sufficiently low values, although this part of the dynamics is, at least approximately, linear. Consequently, the convergence of the method can be accelerated by explicitly exploiting the linearized dynamics in the vicinity of the attracting fixed point. A straightforward method reducing the computational costs by approximately $50\%$ is discussed in \Cref{sec:acceleration}. More sophisticated optimizations can be implemented and will be helpful when applying the adjoint-based variational method to three-dimensional fluid flows. 
    
    \begin{figure}
        \centering
        \includegraphics[width=1.0 \linewidth]{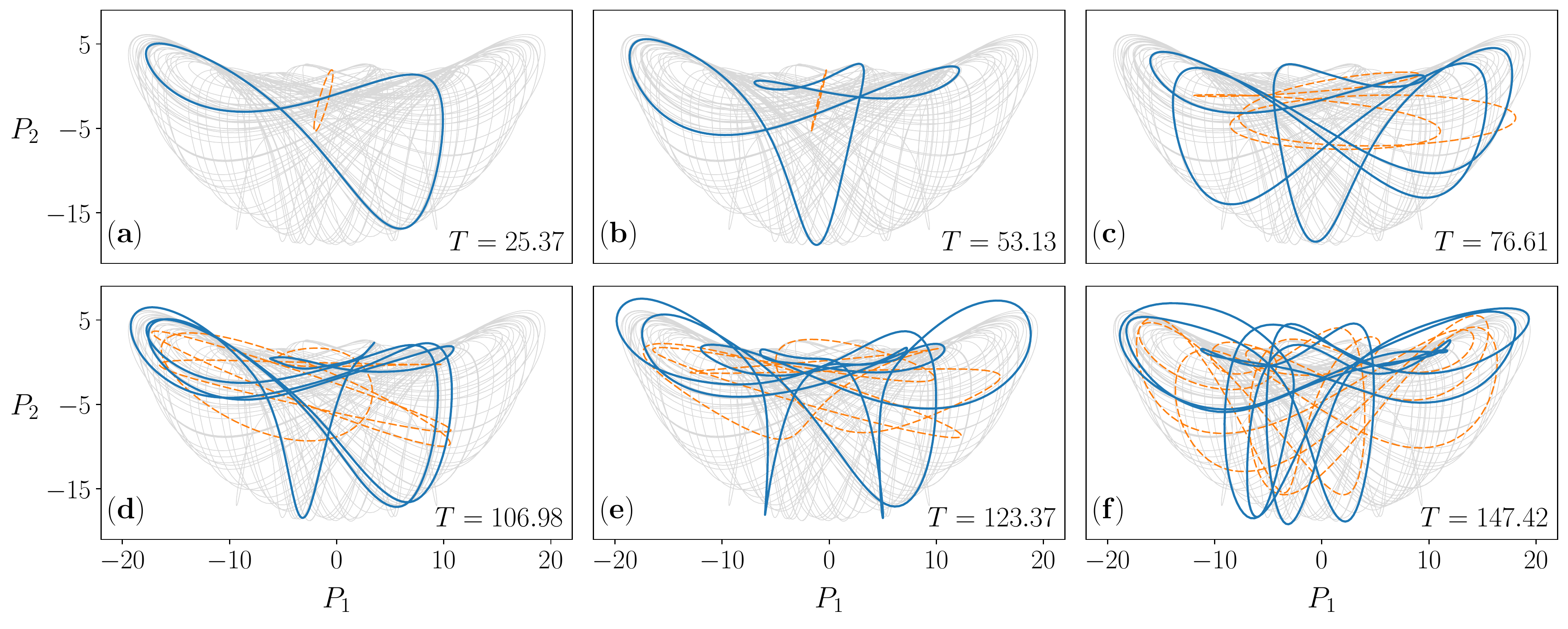}
        \caption{
        Periodic orbits of increasing length and complexity converged by the adjoint-based variational method. The two-dimensional projection of state space as in \Cref{fig:Norm_Loops} indicates, the initial loops (dashed orange lines) as well as the converged periodic orbits (solid blue lines). The period of the converged orbits are given in each panel. The gray line in the background of each panel is the trajectory of a long chaotic solution in the center symmetry subspace of the KSE \cref{eq:KS_sym}. All initial loops are chosen from the center-symmetric subspace. The dynamical system in loop-space preserves the discrete symmetry of the initial loops $\Xi$ so that all converged periodic orbits are also center-symmetric although the symmetry has not been imposed by the method. Note the large differences between initial loops and converged periodic orbits highlighting the global convergence of the adjoint-based variational method.   
        }
        \label{fig:Complex_Phase_Portraits}
    \end{figure}
    
    \begin{figure}
        \centering
        \includegraphics[width=0.7 \linewidth]{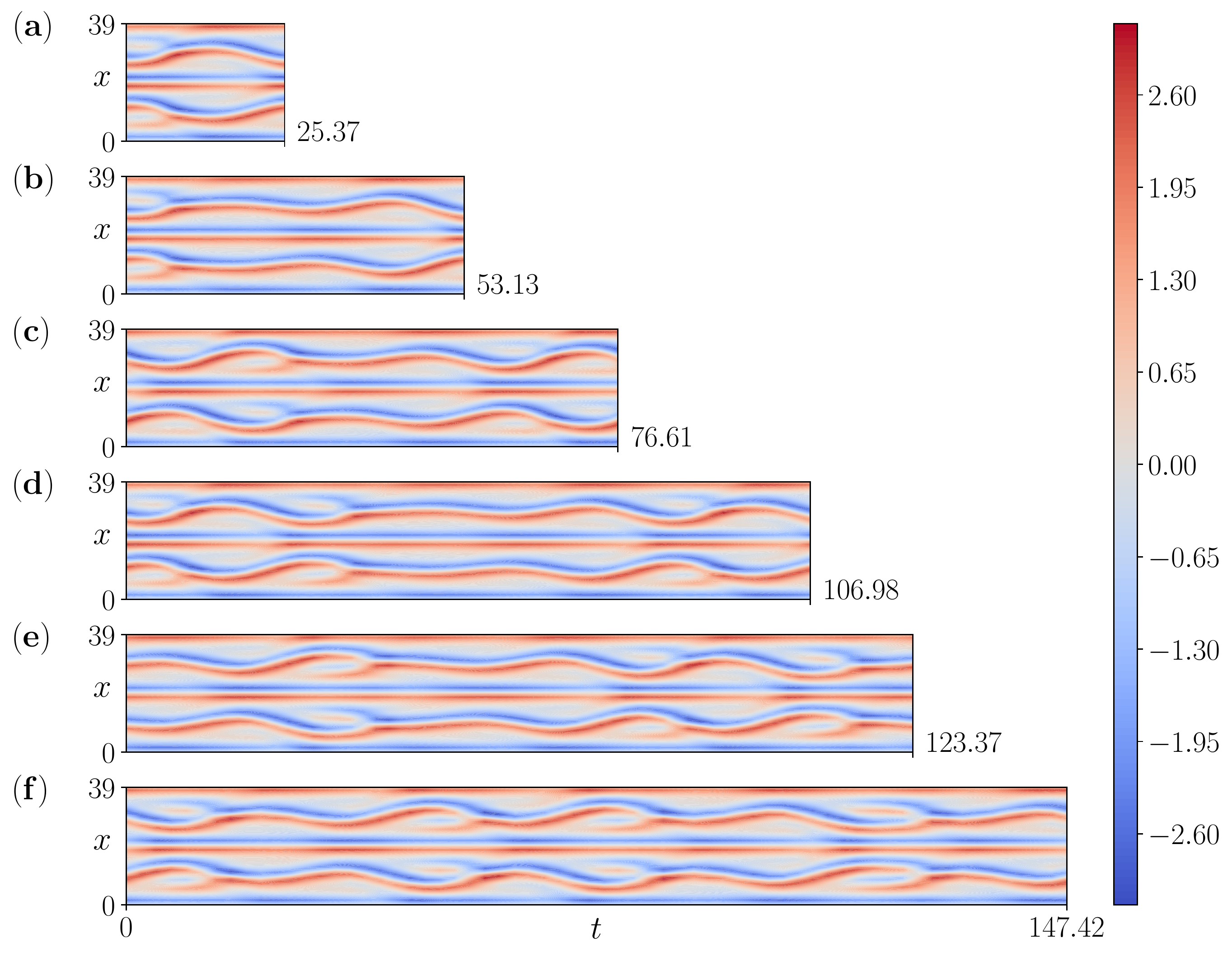}
        \caption{Space-time contours of the converged periodic orbits from \Cref{fig:Complex_Phase_Portraits} with time periods of (a) $T=25.37$, (b) $T=53.13$, (c) $T=76.61$, (d) $T=106.98$, (e) $T=123.37$, and (f) $T=147.42$. Unlike shooting methods, where exponential error amplification during time-integration along the orbit renders long orbits inaccessible, the adjoint-based variational method deforms orbits locally and thus converges independent of the orbit period.}
        \label{fig:Complex_Contours}
    \end{figure}
    
    One major advantage of the adjoint-based variational method is that the successful convergence towards a periodic orbit is independent of the period of the respective orbit. This is in contrast to shooting methods, where the exponential amplification of errors during time-marching along the orbit can hinder computing long orbits. We demonstrate the convergence of orbits of increasing period and complexity in \Cref{fig:Complex_Phase_Portraits}. Six converged periodic orbits with periods ranging from $T=25.37$ to $T=147.42$ are shown in terms of state-space projections, together with initial loops extracted from a chaotic time-series of the KSE. 
    The apparent large difference between initial loop and converged orbit demonstrates that the adjoint-based variational method offers a very large radius of convergence and convergence therefore does not depend on an initial condition in the close vicinity of the converged orbit. 
    The evolution of loops under the dynamical system in loop-space converges to minima of the cost function $J$ for any initial condition. While globally convergent, the variational method is not guaranteed to converge to absolute minima of $J$ with $J=0$, corresponding to periodic orbits, but the dynamics may approach a local minimum with $J>0$. For initial loops extracted from recurrences in a one-dimensional projection of state space, as discussed in \ref{sec:initial_converge}, we observe approximately $70\%$ of all initial conditions to converge to periodic orbits with $J=0$. An example of a loop approaching a local minimum of $J$ is shown in \Cref{sec:app_conv_ex}.
    
    Following Lasagna \cite{Lasagna2017}, initial loops for the six orbits discussed in \Cref{fig:Complex_Phase_Portraits} are extracted from a chaotic trajectory of the KSE in the subspace of center-symmetric fields. All initial conditions for the initial value problem in loop-space are therefore center symmetric. The dynamical system in loop-space \cref{eq:KS_G} preserves the symmetry $\Xi$ of loops \cref{eq:KS_loop_sym} that corresponds to the center symmetry of instantaneous fields in the KSE system \cref{eq:KS_sym}.
    Consequently, all converged periodic orbits also lie in the center symmetry subspace, as confirmed by \Cref{fig:Complex_Contours}, where space-time contours of the six periodic orbits are shown. Note that the method does not explicitly enforce the discrete symmetry but preserves the symmetry of the initial condition. 

\section{Summary and conclusion}
\label{sec:conclusion}
    Unstable periodic orbits have been recognized as building blocks of the dynamics in driven dissipative spatio-temporally chaotic systems including fluid turbulence. Periodic orbits capture key features of the dynamics and reveal physical processes sustaining the turbulent flow. Constructing a sufficiently large set of periodic orbits moreover carries the hope to eventually yield a predictive rational theory of turbulence, where `properties of the turbulent flow can be mathematically deduced from the fundamental equations of hydrodynamics', as expressed by Hopf in 1948 \cite{Hopf1948}.
    Despite the importance of unstable periodic orbits, computing these exact solutions for high-dimensional spatio-temporally chaotic systems remains challenging. Known methods either show poor convergence properties because they are based on time-marching a chaotic system causing exponential error amplification; or they require constructing Jacobian matrices which is prohibitively expensive for high-dimensional problems. We therefore introduce a new matrix-free method for computing periodic orbits that is unaffected by exponential error amplification, shows robust convergence properties and can be applied to high-dimensional spatio-temporally chaotic systems. As a proof-of-concept we implement the method for the one-dimensional KSE and demonstrate the convergence of periodic orbits underlying spatio-temporal chaos.
    
    The adjoint-based variational method constructs a dynamical system that evolves entire loops such that the value of a cost function measuring deviations of the loop from a solution of the governing equations monotonically decreases. Periodic orbits correspond to attracting fixed points of the variational dynamics.
    Due to the variational approach, the method provides a large radius of convergence so that periodic orbits can be found from inaccurate initial guesses. For the KSE we demonstrate the robust convergence properties by successfully computing periodic orbits from inaccurate initial guesses. These guesses are extracted from the projection of the free chaotic dynamics on a single scalar quantity, instead from close recurrences based on the $L_2$-distance between spatial fields \cite{Auerbach1987}. Reliable convergence to machine precision is observed independent of the period of the orbit.
    
    The large convergence radius of the adjoint-based variational method relaxes accuracy requirements for initial guesses when those are extracted from the chaotic dynamics. Since initial guesses are characterized by an entire loop, one may use fast-to-compute models approximating the full dynamics to construct initial guesses for periodic orbits of the full dynamics. Such an approach would not be reasonable for classical shooting methods where initial guesses are characterized by an instantaneous initial condition and the difference between model and full dynamics would be amplified exponentially by the time-marching. Suitable models that may help provide initial guesses for constructing large sets of periodic orbits for a given chaotic system include under-resolved simulations, spatially filtered equations such as LES in fluids applications \cite{Sagaut2006} and classical POD / DMD based models \cite{McKeon2017}. In addition, recent breakthroughs in machine learning allow to create data-driven low-dimensional models of the chaotic dynamics that replicate spatio-temporal chaos in one- and two-dimensional systems with remarkable accuracy \cite{Pathak2018,Zimmermann2018,Vlachas2018}.
    
    The feasibility of the proposed method has been demonstrated for a one-dimensional chaotic PDE but the method applies to general autonomous systems and we plan to implement it for the full three-dimensional Navier-Stokes equations. Specifically, we aim for an implementation within our own open-source software Channelflow (\href{https://www.channelflow.ch}{channelflow.ch}) \cite{gibson2019}. In the context of this software not only the identification of periodic orbits but also their numerical continuation will benefit from the adjoint-based variational approach. When transferring the adjoint-based variational approach to three-dimensional fluid turbulence, we envision further optimizations of the method. 
    First, we will exploit that during its approach to the attracting fixed point representing the periodic orbit, the evolution is well approximated by the linearization of the dynamics around the attracting fixed point. This allows to accelerate the time-marching in loop-space and thereby the exponential convergence, as exemplified for the KSE. Second, one may complement the adjoint dynamics with Newton descent to identify the attracting fixed point in loop-space, following the analogous hybrid approach for identifying equilibrium solutions \cite{Farazmand2016}. Alternatively, we will combine the adjoint-based variational method with a Newton-GMRES-based shooting method. Such a hybrid method offers the large radius of convergence of the adjoint-based variational method in combination with the fast quadratic convergence of Newton's method. To allow for converging long and unstable periodic orbits, a multi-shooting variant of the standard Newton-GMRES-hook-step method \cite{Sanchez2010} will be used.

\appendix
\section{Rate of change of the cost function $J$}
\label{sec:d(J)/d(tau)}
    The rate of change of the cost function $J$ with respect to the fictitious time $\tau$ is given in \Cref{eq:rate_of_J}. Here we derive this expression including the specific form of $\glp{\resv}$. 
    With the definition of the cost function $J$ \cref{eq:cost_function}
    \begin{align*}
        J(\lp)=\int_0^1\int_\Omega{\vec{\res}(\lp).\vec{\res}(\lp)}d\vec{x}ds,
    \end{align*}
    the rate of change of $J$ with respect to the fictitious time $\tau$ is
    \begin{align*}
        \frac{\partial J}{\partial \tau}=2\int_0^1\int_\Omega{\left(\nabla_\lp\vec{\res}\cdot\glp{G}\right)\cdot \vec{\res}}d\Vec{x}ds.
    \end{align*}
    where $\partial\glp{l}/\partial\tau=\glp{G}$ from definition \cref{eq:lp_PDE} has been used. 
    Using the definition of the inner product in the space of generalized loops \cref{eq:inner_product_definition}, we can express the rate of change as
    \begin{align*}
    \renewcommand\arraystretch{1.5}
        \frac{\partial J}{\partial \tau}=
        2\left<\;
        \begin{bmatrix}
            \nabla_\lp \vec{\res}\cdot\glp{G}\\
            0
        \end{bmatrix}
        ,
        \begin{bmatrix}
            \vec{\res}\\
            0
        \end{bmatrix}
        \;\right>.
    \renewcommand\arraystretch{1}
    \end{align*}
    Here we choose the second component of both generalized loops to be zero. With this choice, the rate of change of $J$ is given by
    \begin{align*}
        \frac{\partial J}{\partial \tau}=2\left<\ddrv(\glp{l};\glp{G}),\glp{\resv}\right>,
    \end{align*}
    where $\ddrv(\glp{l};\glp{G})$ indicates the directional derivative of $\glp{\resv}=[\vec{\res};0]$ along $\textbf{G}$, defined in \cref{eq:directional_derivative}.

\section{Adjoint operator for KSE}
\label{sec:K-S_adjoint_operator}
    We explicitly derive the form of the adjoint operator for the KSE problem given in \Cref{eq:KS_L_dagger}. 
    In this appendix, subscripts 1 and 2 denote the field component and the scalar component of generalized loops, respectively. The directional derivative of KSE along $\glp{G}$ is
    \begin{align*}
    \renewcommand\arraystretch{1.8}
        \ddrv(\lp;\glp{G})=
        \begin{bmatrix}
            \dfrac{G_2}{T^2}\dfrac{\partial u}{\partial s}-\dfrac{1}{T}\dfrac{\partial G_1}{\partial s}-\dfrac{\partial (uG_1)}{\partial x}-\dfrac{\partial^2 G_1}{\partial x^2}-\dfrac{\partial^4 G_1}{\partial x^4}\\
            0
        \end{bmatrix}
    \end{align*}
    To compute the adjoint operator, we expand the inner product of the directional derivative of the residual and the residual itself:
    \begin{align}
         &\left<\ddrv(\lp;\glp{G}),\glp{\resv}\right> \nonumber\\
         &\quad\quad=\int_0^1\int_0^L{\ddr_1\resv_1 dxds}+\ddr_2\resv_2=\int_0^1\int_0^L{\ddr_1\resv_1 dxds}+0 \nonumber\\
         &\quad\quad=\int_0^1\int_0^L{\left(\frac{G_2}{T^2}\frac{\partial u}{\partial s}-\dfrac{1}{T}\frac{\partial G_1}{\partial s}-\frac{\partial (uG_1)}{\partial x}-\frac{\partial^2 G_1}{\partial x^2}-\frac{\partial^4 G_1}{\partial x^4}\right)\resv_1}dxds\nonumber\\
         &\quad\quad=\int_0^1\int_0^L{\frac{G_2}{T^2}\frac{\partial u}{\partial s}\resv_1}dxds \label{eq:app_directional}\\
         &\quad\quad\quad+\int_0^1\int_0^L{\left(-\dfrac{1}{T}\frac{\partial G_1}{\partial s}-\frac{\partial (uG_1)}{\partial x}-\frac{\partial^2 G_1}{\partial x^2}-\frac{\partial^4 G_1}{\partial x^4}\right)\resv_1}dxds.\nonumber
    \end{align}
    This inner product must be equal to
    \begin{equation}\label{eq:app_adjoint}
        \left<\glp{G},\adjv(\lp;\glp{\resv})\right> = \int_0^1\int_0^L{\adj_1G_1dxds}+\adj_2G_2,
    \end{equation}
    where the adjoint operator is indicated by a dagger. 
    Direct comparison of equations \cref{eq:app_directional} and \cref{eq:app_adjoint} results in
    \begin{subequations}
        \begin{align}
            \int_0^1\int_0^L{\adj_1G_1dxds}&=\int_0^1\int_0^L{\left(-\dfrac{1}{T}\frac{\partial G_1}{\partial s}-\frac{\partial (uG_1)}{\partial x}-\frac{\partial^2 G_1}{\partial x^2}-\frac{\partial^4 G_1}{\partial x^4}\right)\resv_1dxds} \label{eq:Adjoint_G1}\\
            \adj_2G_2&=\left(\int_0^1\int_0^L{\frac{1}{T^2}\frac{\partial u}{\partial s}\resv_1}dxds\right)G_2. \label{eq:Adjoint_G2}
        \end{align}
    \end{subequations}
    The form of $\adj_2$ is directly given by \cref{eq:Adjoint_G2}:
    \begin{align*}
        \adj_2(\glp{q};\glp{\resv})=\int_0^1\int_0^L{\frac{1}{T^2}\frac{\partial u}{\partial s}\resv_1}dxds.
    \end{align*}
    Using integration by parts and the periodicity of the domain in space and time, \Cref{eq:Adjoint_G1} becomes
    \begin{align*}
        \int_0^1\int_0^L{\adj_1G_1dxds}=\int_0^1\int_0^L{\left(\dfrac{1}{T}\frac{\partial \resv_1}{\partial s}+u\frac{\partial \resv_1}{\partial x}-\frac{\partial^2 \resv_1}{\partial x^2}-\frac{\partial^4 \resv_1}{\partial x^4}\right)G_1dxds}.
    \end{align*}
    Consequently,
    \begin{align*}
        \adj_1(\lp;\glp{\resv})=\frac{1}{T}\frac{\partial \resv_1}{\partial s}+u\frac{\partial \resv_1}{\partial x}-\frac{\partial^2 \resv_1}{\partial x^2}-\frac{\partial^4 \resv_1}{\partial x^4}
    \end{align*}
    where $\resv_1=\res$. The adjoint operator acting on loops therefore has the form
    \begin{align*}
    \renewcommand\arraystretch{2.5}
        \adjv(\lp;\glp{\resv})=
        \begin{bmatrix}
            \dfrac{1}{T}\dfrac{\partial \res}{\partial s}+u\dfrac{\partial \res}{\partial x}-\dfrac{\partial^2 \res}{\partial x^2}-\dfrac{\partial^4 \res}{\partial x^4}\\
            \displaystyle\int_0^1\int_0^L{\frac{1}{T^2}\frac{\partial u}{\partial s}\res}dxds
        \end{bmatrix}.
    \end{align*}

\section{Acceleration of the convergence by linearized approximation}
\label{sec:acceleration}
    \begin{figure}
        \centering
        \includegraphics[width=0.6 \linewidth]{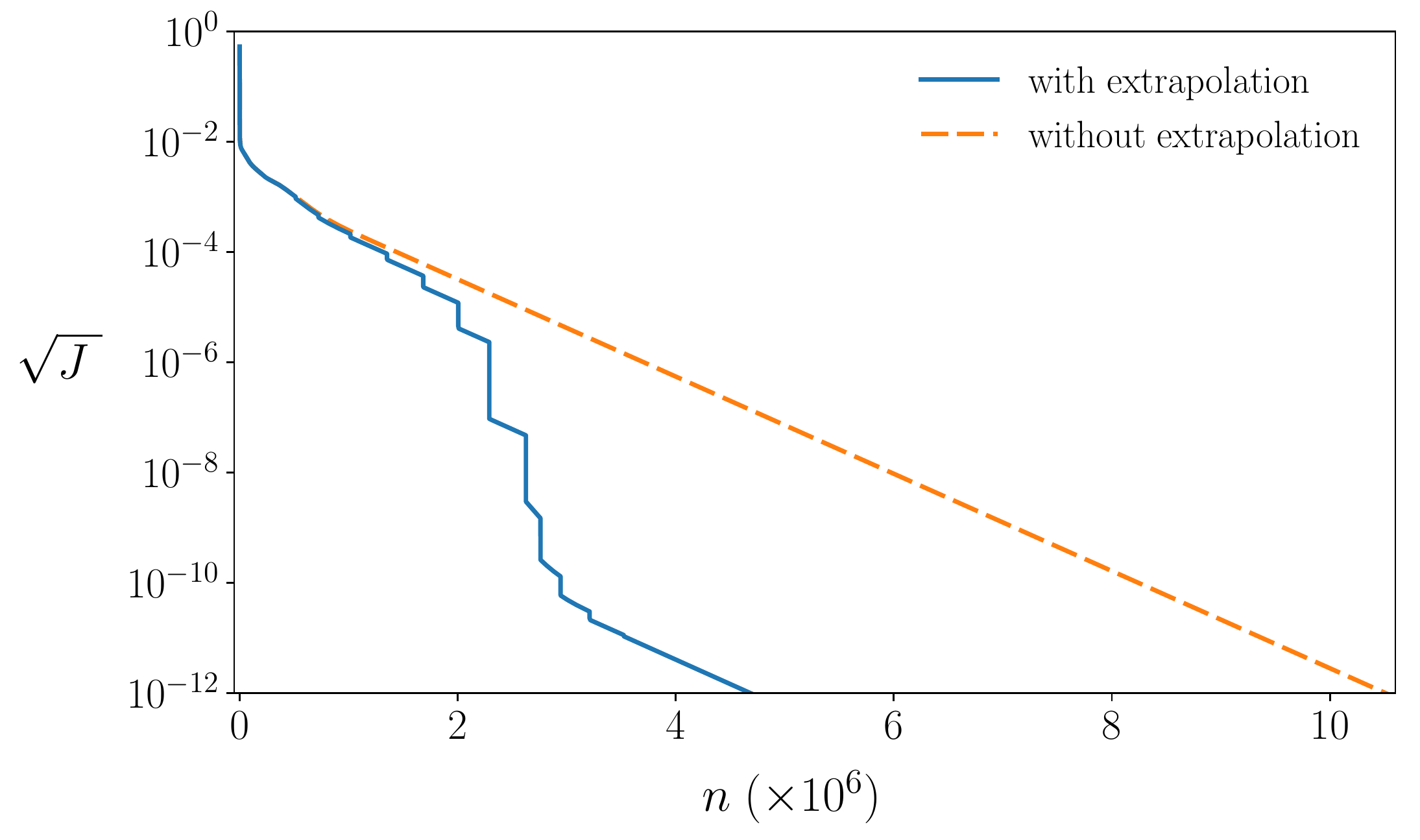}
        \caption{Accelerated convergence of the adjoint-based variational method. Convergence history for the periodic orbit discussed in figures \ref{fig:Norm_Loops} and \ref{fig:T_Contour}, for the standard method (orange dashed line) and the modified method involving linear extrapolations along the solution trajectory in the loop-space.
        The linear extrapolations are based on a linear approximation of the loop dynamics around the attracting fixed point in loop-space corresponding to the periodic orbit.
        The square root of the cost function is shown as a function of the number of fictitious time steps $n$. The first extrapolation is performed when $\sqrt{J}=10^{-3}$. Between two consecutive extrapolations, the dynamical system in loop-space is integrated until the value $\sqrt{J}$ is halved. 
        In this example case, extrapolations reduce the total number of fictitious time steps by more than $50\%$.
        }
        \label{fig:Extrapolation}
    \end{figure}
    
    We demonstrate a straightforward method for accelerating the convergence of the adjoint-based variational method. We iterate between time-stepping of the dynamical system in loop-space \cref{eq:KS_G} and a linear extrapolation along the evolution trajectory of the loops. This extrapolation is based on the assumption that the evolution follows the leading eigendirection of the linearization about the attracting loop. Extrapolations yield the initial conditions of the subsequent advancing of the loop in $\tau$. This procedure is repeated until the periodic orbit is converged. \Cref{fig:Extrapolation} compares the convergence of the periodic orbit shown in figures \ref{fig:Norm_Loops} and \ref{fig:T_Contour} by continuous integration of the dynamical system in loop-space \cref{eq:KS_G} and the accelerated method iterating between time-stepping of the full dynamics and extrapolations, both from the same initial condition. Vertical drops of the cost function shown in the graph correspond to the extrapolations. In this example the accelerated method reduces the required total number of numerical steps of integration by more than $50\%$.

\section{Convergence to local and global minima of $J$}
\label{sec:app_conv_ex}
    Here we show an example of time-stepping of the dynamical system in loop-space where the final loop corresponds to local minimum of $J$ with a nonzero value. Consequently, no periodic orbit is found.
    
    \begin{figure}
        \centering
        \includegraphics[width=0.85 \linewidth]{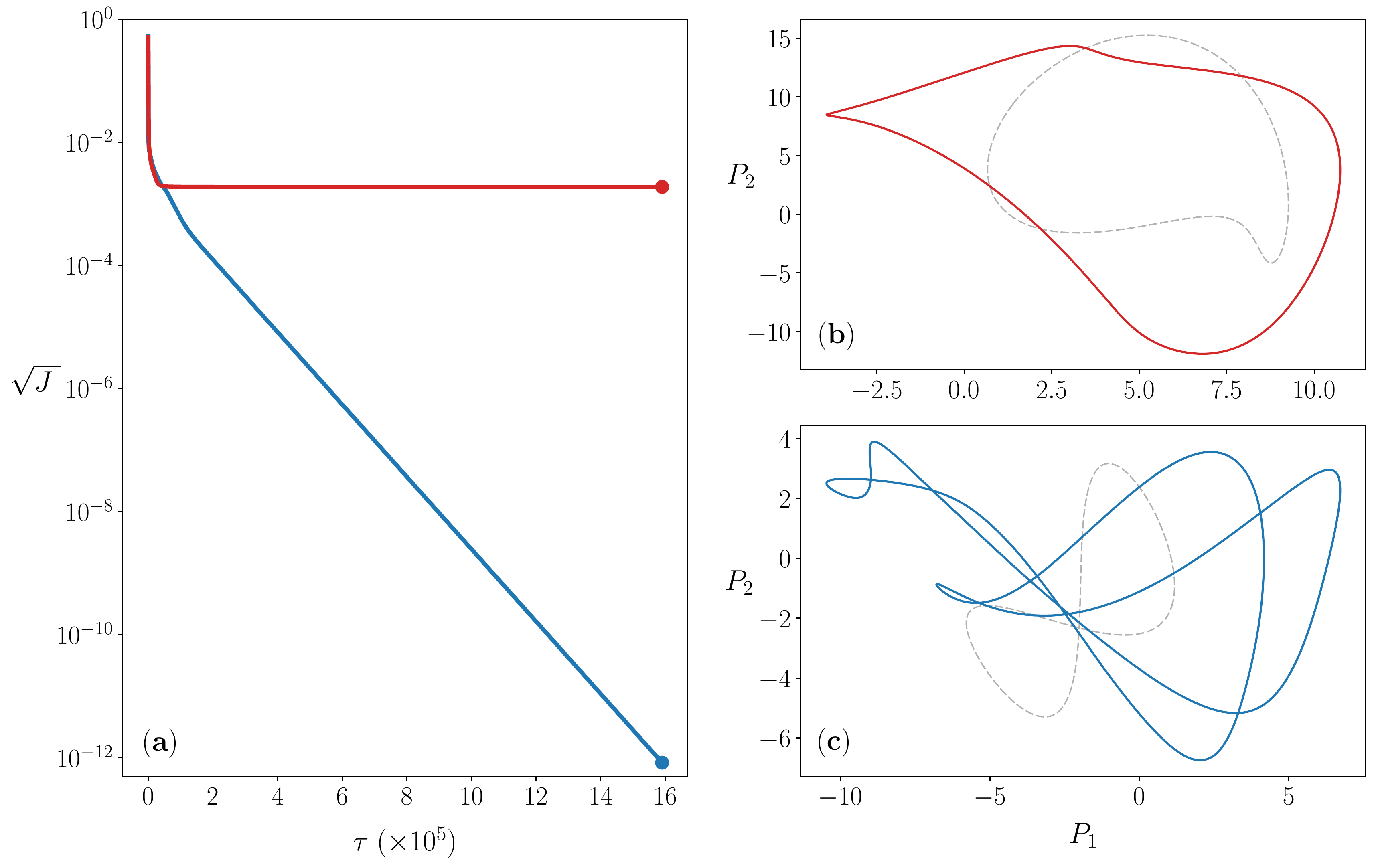}
        \caption{Minimizing $J$ by the adjoint-based variational method.
        $(a)$ Evolution of $\sqrt{J}$ with $\tau$ for two different initial loops.
        The blue line shows the convergence for the loop that approaches a periodic orbit with $J=0$ while the red line shows the convergence for a loop that approaches a local minimum of $J$ with a nonzero value $J > 0$. The corresponding initial (dashed lines) and converged loops (solid lines) in the two-dimensional projection of the state space as in \Cref{fig:Norm_Loops} are visualized for the converged loop with $J > 0$ (panel b) and the periodic orbit with $J \rightarrow 0$ (panel c).
        }
        \label{fig:Failed_vs_Successful}
    \end{figure}
    
\section*{Acknowledgments}
    We thank Florian Reetz for insightful discussions on the implementation of the proposed method both for the KSE but also for future implementations within Channelflow. SA acknowledges support by the State Secretariat for Education, Research and Innovation SERI via the Swiss Government Excellence Scholarship.
    
\bibliographystyle{siamplain}
\bibliography{references}
\end{document}